\documentstyle[sprocl]{article}

\def\be{\begin{equation}}
\def\ee{\end{equation}}
\def\bea{\begin{eqnarray}}
\def\eea{\end{eqnarray}}

\begin{document}
\pagestyle{empty}

\title{MAGNETIC CATALYSIS OF CHIRAL SYMMETRY BREAKING AND THE PAULI PROBLEM}

\author{ Y. JACK NG}

\address{Institute of Field Physics, Department of Physics and 
Astronomy, \\ University of North Carolina, Chapel Hill, NC  27599-3255, 
USA\\E-mail: ng@physics.unc.edu}

\maketitle\abstracts 
{ The non-perturbative Schwinger-Dyson equation is used to show that chiral
symmetry is dynamically broken in QED at weak gauge couplings when an
external uniform magnetic field is present.  A complete analysis of this
phenomenon may shed light on the Pauli problem, namely, why $\alpha$ = 1/137.}

Let me begin with a joke which some of you may have heard before.  One
of Wolfgang Pauli's life-long dreams was to understand why the fine
structure constant in electrodynamics is 1/137 (in the infrared regime).  
Pauli was also known to be a difficult person, very hard to
please.  As the joke goes, the first thing Pauli
asked God after his death was to explain why $\alpha$ = 1/137.  As God went 
on with His
explanation, Pauli grew more and more dissatisfied.  After five minutes,
Pauli was seen storming out of Heaven's Gates mumbling, "Ridiculous!"

Like Pauli, I also would like to understand why $\alpha$ = 1/137.  To
dignify this problem, I will call it the Pauli problem.  It is possible
that chiral symmetry breaking by an external field in QED may 
provide some insight
on this old problem by giving a critical value of $\alpha$ close to
1/137.\cite{JBW}  Admittedly, nothing close to that magic value has 
arisen in the
results we have obtained so far \cite{Lee}, but our analysis is not yet
complete.

My interest in chiral symmetry breaking by an external
field dates back a dozen years ago, to the time when a
multiple correlated and narrow-peak structures in electron and 
positron spectra\cite{GSI} was observed 
in heavy-ion experiments at GSI.  Kikuchi and I
\cite{newphase} interpreted the $e^+e^-$ peaks as decay products of a new
type of positronium, which is formed in a new QED phase 
induced by the electromagnetic fields of the colliding heavy ions.  The
theoretical underpinning of this scenario was provided by earlier works
\cite{Mir} which indicated that QED might have a non-perturbative
strong-coupling phase, characterized by spontaneous chiral symmetry
breaking, in addition to the familar weak-coupling phase.  The negative
results in recent heavy-ion collision experiments at Argonne
\cite{Argonne} have rendered our interpretation moot.  Nevertheless,
the problem of chiral symmetry breaking by an external field is still
interesting as it may shed light on the Pauli problem, and as it
provides an example of vacuum engineering by manipulating external fields
to alter the symmetry properties of the vacuum. But more concretely, our
study provides a new non-perturbative phenomenon in (3+1)-dimensional
quantum field theories and a new method to study it.

First, what kind of external fields can induce chiral symmetry breaking 
in gauge theories?\cite{Ng}  Lessons gained from studying the 
Nambu-Jona-Lasinio model \cite{KL} lead us to believe that uniform
magnetic fields
are prime candidates.  To put our problem in as general a setting as
possible, we want an approach that treat both the coupling and the
external field non-perturbatively.  The former criterion is met by
using the Schwinger-Dyson equations (or equivalently, the
Nambu-Bethe-Salpeter equations \cite{GMS}); the latter condition is
satisfied by applying the strong-field techniques introduced by Schwinger
and others.

Let us start with the motion of a massless fermion of charge $e$ 
in an external 
electromagnetic field.  It is described by the Green's function that 
satisfies the modified Dirac equation proposed by Schwinger:
\begin{equation}
\gamma \cdot \Pi(x) G_A(x,y) + \int d^4x' M(x,x') G_A(x',y) = 
\delta^{(4)}(x-y),
\label{Greeneq}
\end{equation}
where $\Pi_\mu(x) = - i \partial_\mu - e A_\mu(x)$, and $M(x,x')$ 
is the mass operator $M$ in the coordinate representation.  For a 
constant magnetic field of strength $H$, we may take $A_2 = Hx_1$ 
to be the only nonzero component of $A_\mu$.  We adopt 
the method due to Ritus\cite{Ritus}, which is based on the 
use of the eigenfunctions of the mass operator and the diagonalization 
of the latter.  As shown by Ritus, $M$ is diagonal in the 
representation of the eigenfunctions $E_p(x)$ of the operator 
$(\gamma \cdot \Pi)^2$:  
\begin{equation}
- (\gamma \cdot \Pi)^2  E_p(x) = p^2 E_p(x).
\label{eigeneq}
\end{equation}
The advantage of using this representation is obvious: $M$ can now 
be put in terms of its eigenvalues, so the problems arising from its 
dependence on the operator $\Pi$ can be avoided.  In the chiral 
representation in which $\sigma_3$ and $\gamma_5$ are diagonal 
with eigenvalues $\sigma = \pm 1$ and $\chi = \pm 1$, respectively, 
the eigenfunctions $E_{p\sigma\chi}(x)$ take the form
\begin{equation}
E_{p\sigma\chi}(x) = N {\rm e}^{i (p_0x^0 + p_2x^2 + p_3x^3)} D_n(\rho) 
\omega_{\sigma\chi} \equiv \tilde{E}_{p\sigma\chi} \omega_{\sigma\chi},
\label{eigenfcn}
\end{equation}
where $D_n(\rho)$ are the parabolic cylinder functions
with indices 
\begin{equation}
n = n(k,\sigma) \equiv k + \frac{e H \sigma}{2 |e H|} - \frac{1}{2},
~~~~k = 0, 1, 2, ...,
\label{index}
\end{equation}         
and argument $\rho = \sqrt{2 |e H|} (x_1 - \frac{p_2}{e H})$.  
Note that $n = 0,~1,~2,~...~$.  
The normalization factor is $N = (4 \pi |eH|)^{1/4}/\sqrt{n!}$; 
$p$ stands for the set $(p_0, p_2, p_3, k)$; and 
$\omega_{\sigma\chi}$ are the bispinors of $\sigma_3$ and 
$\gamma_5$.

Following Ritus, we form the orthonormal and complete 
eigenfunction-matrices $E_p = {\rm diag}(\tilde{E}_{p11},~
\tilde{E}_{p-11},~\tilde{E}_{p1-1},~\tilde{E}_{p-1-1})$.  They 
satisfy 
\begin{equation}
\gamma \cdot \Pi~E_p(x) = E_p(x)~\gamma \cdot \bar{p}
\end{equation}
and 
\begin{equation}
M(x,x') E_p(x') = E_p(x) \delta^{(4)}(x-x') {\Sigma}_A(\bar{p}),
\label{masseigeneq}
\end{equation}
where ${\Sigma}_A(\bar{p})$ represents the eigenvalues of 
the mass operator, and $\bar{p}_0 = p_0,~\bar{p}_1 = 0,~\bar{p}_2 
= - {\rm sgn}(eH) \sqrt{2|eH|k},~\bar{p}_3 = p_3$.  These 
properties of the $E_p(x)$ allow us to express the Green's function 
in the $E_p$-representation as 
$(\bar{E}_p \equiv \gamma^0 E_p^\dagger \gamma^0)$
\begin{equation}
G_A(x,y) = \Sigma \!\!\!\!\!\! \int \frac{d^4p}{(2 \pi)^4} E_p(x) \frac{1}
{\gamma \cdot \bar{p} + {\Sigma}_A(\bar{p})} \bar{E}_p(y),
~~~\Sigma \!\!\!\!\!\! \int d^4p \equiv \sum_{k} \int dp_0 dp_2 dp_3.
\label{Greenfcn}
\end{equation}

We work in the ladder quenched approximation. 
In terms of the notations: 
$\bar{p"}_{\!\!\!\!_0} = p_0 - q_0$, $\bar{p"}_{\!\!\!\!_1} = 0$, 
$\bar{p"}_{\!\!\!\!_2} = -~{\rm sgn}(eH) \sqrt{2|eH|k"}$, 
$\bar{p"}_{\!\!\!\!_3} = p_3 - q_3$, 
the Schwinger-Dyson equation takes the form 
\begin{equation}
\Sigma_A(\bar{p}) \simeq \frac{i e^2}{(2 \pi)^3} |eH| \int dq_0 dq_3 
\int_0^\infty dr^2 {\rm e}^{- r^2} \frac{-2}{q^2} \frac
{\Sigma_A(\bar{p"})}{\bar{p"}^2 + \Sigma_A(\bar{p"})}
\label{fermass}
\end{equation}
where $q^2 = - q_0^2 + q_3^2 + 2 |eH| r^2$ and $\bar{p"}^2 = 
- (p_0 - q_0)^2 + (p_3 - q_3)^2 + 2 |eH| k"$, and where summation over
$k'' = k, ~k \pm 1$ is understood.  We have assumed that, due to the
factor e$^{- r^2}$ in the integrand, contributions from large values of $r$ 
are suppressed.

Let us make a Wick rotation to Euclidean space: $p_0 \rightarrow i p_4$, 
$q_0 \rightarrow i q_4$, and consider the case with $p = 0$, i.e., $p_0 = p_3 
= k = 0$.  We assume that the dominant 
contributions to the integral in Eq.(\ref{fermass}) come from the 
infrared region of small $q_3$ and $q_4$, and that the $k'' = 0$ term dominates
over the $k'' = 1$ term (this assumption has been checked 
to be self-consistent).  Then, it is reasonable to replace 
$\Sigma_A(\bar{p"})$ in the integrand by $\Sigma_A(0) = m \times 
{\bf 1}$.  The SD equation yields the nonzero dynamical mass as
\begin{equation}
m \simeq a~\sqrt{|eH|}~{\rm e}^{- b\sqrt{\frac{\pi}{\alpha}}},
\label{result}
\end{equation} 
where $a$ and $b$ are constants of order 1.

Eq.(\ref{result}) clearly demonstrates the nonperturbative nature of 
the result.  
It also shows that our approximations break down when 
$\alpha > O(1)$.  
Our earlier 
assumption that effectively $r \ll 1$ is translated to the 
physical assumption that $m/\sqrt{|eH|}  \ll 1$, which requires that 
$\alpha \ll O(1)$; in other words, the dynamical chiral 
symmetry breaking solution we have found applies to the 
weak-coupling regime of QED!  We have checked that indeed 
the infrared region of $q_3$ and $q_4$ gives the dominant 
contributions to the integrals.

To establish that the above solution to the SD equation for the fermion
self-energy does indeed correspond to a dynamical chiral symmetry-breaking
solution, it is necessary to demonstrate the existence of the
corresponding Nambu-Goldstone boson.  One way to establish this is by
studying the Nambu-Bethe-Salpeter equation of the bound-state NG boson, as
was done by Gusynin {\it et al.} \cite{GMS}, who found a solution
consistent with our Eq.(\ref{result}).  As a consistency check on our
approach, we \cite{Lee} have recovered the same result from the NBS
equation, using the $E_p$-representation of the fermion propagator.  The
chiral condensate, the order parameter for dynamical symmetry breaking,
can be easily computed, based on our formalism.  We obtain ($\psi$ is the
fermion field)
\begin{equation}
\langle \bar{\psi} \psi \rangle
~\simeq~
-~\frac{|eH|}{2 \pi^2} ~m ~\ln\left(\frac{|eH|}{m^2}\right).
\label{psibpsi}
\end{equation}

Thermal effects and the effects of a chemical potential can be readily
incorporated into our study of chiral symmetry breaking in an external
magnetic field. \cite{Lee}  One finds that chiral symmetry is restored
above a critical temperature which is of the order of the dynamical
fermion mass (given by Eq.(\ref{result})), and the corresponding phase
transition is of second order.  In contrast, the chiral symmetry
restoration above a critical chemical potential (also of the order of the
dynamical fermion mass) is a first order phase transition.  The
chiral condensates for both cases have the same form as given above in
Eq.(\ref{psibpsi}).

To summarize, we have shown that
chiral symmetry is dynamically broken in quenched, ladder QED when an
external magnetic field is present.  So, the external magnetic field acts as
a catalysis for chiral symmetry breaking.
Furthermore, in the lowest Landau 
level approximation, this chiral symmetry
breaking is generated in the infrared region where the QED gauge coupling
is weak.  As pointed out by Gusynin {\it et al.} \cite{GMS}, the dynamics
of LLL is (4 minus 2)-dimensional [see the second propagator factor with 
$k'' = 0$ in
Eq.(\ref{fermass})].  But this effective dimensional reduction is only for
charged states.  Propagators for neutral states (like the photon and NG
boson) are still 4-dimensional [see the first propagator factor in
Eq.(\ref{fermass})].  Thus the Mermin-Wagner-Coleman theorem (which
stipulates that there can be no spontaneous breakdown of continuous
symmetries in dimensions less than three) is successfully evaded by our
chiral symmetry breaking solution.  In fact, it is the interplay
between the 2-dimensional dynamics of the charged particles (in the LLL 
approxiamtion) and the
4-dimensional dynamics of the neutral particles that is partly
responsible for some of the characteristics of our chiral
symmetry-breaking solution.  
For the magnetic catalysis to take place, the Landau
energy $(\sqrt{|eH|})$ must be much greater than the fermion mass.  If the
fermion mass is taken to be zero, the critical magnetic field is zero.  
Thus chiral symmetry is broken at an arbitrarily weak magnetic field.  In
the LLL approximation, the fermion pairing dynamics responsible for chiral
condensates is (1 + 1)-dimensional, and is hence strong in the infrared
region.  Thus chiral symmetry breaking takes place even at the weakest
attractive interactions between the fermions, yielding a zero critical
gauge coupling.

As for applications, chiral symmetry breaking by an external 
magnetic field may be relevant to condensed matter systems (e.g., in
quantum Hall effect and in superconductivity \cite{KM}).  It may also find
applications in astrophysics (e.g., in the cooling process of neutron
stars).  It has been suggested in the literature \cite{GMS} that the
chiral symmetry-breaking solution may play a role in the electroweak
phase transition during the early evolution of the Universe since a huge
magnetic field (with the Landau energy of the order of electroweak
breaking scale) was purportedly generated by the phase transition.
Unfortunately, the electroweak phase transition took place at a
temperature also of the order of the electroweak scale, 
which is much higher than the critical
temperature for a magnetic field of such a magnitude.  Therefore, it is very
unlikely that such a magnetic field could change the character of the
electroweak phase transition in any way.  But, if (for reasons not yet
understood) magnetic fields of such extraordinarily large magnitudes as 
to overwhelm the thermal effects were generated in the early epoch, then
the chiral symmetry-breaking solution reported here could be very relevant
to the early evolution of the Universe.

There remain several interesting questions which we intend to investigate.
For instance, how do other background field configurations affect chiral
symmetry breaking?  (As an example, it is expected that an electric field
would tend to break up the condensate and destabilize the vacuum, thus
inhibiting chiral symmetry breaking \cite{DW}).  Are there strong-coupling
solutions of chiral symmetry breaking in an external magnetic field?  If
so, are the weak-coupling solutions we have found and the new
strong-coupling solutions related by some kind of duality?  How does one 
extend the chiral symmetry-breaking results to inhomogenous
configurations?  Can one understand the chiral symmetry breaking by a
simple (perhaps topological) argument? \cite{Sem}

But, perhaps the most urgent task is to have a better
understanding of this phenomenon of
chiral symmetry breaking by a magnetic field.  After all,
if dynamical
symmetry breaking occurs when particles interact strongly as in QCD, then
how can {\it weak} electromagnetic interactions in an arbitrarily {\it
weak} magnetic field lead to chiral symmetry breaking?   Above, we 
have given a 
partial answer to this question by arguing that the dimensional reduction
in the LLL approximation effectively enhances such weak interactions in
the infrared region.  But can the LLL approximation be the whole story?

For a better perspective, it is useful to
recall that in strong-coupling QED
\cite{Mir}, where chiral symmetry is broken, the anomalous dimension of the
fermion mass operator is one, thus four-fermion interactions are relevant
operators.  In fact, it is the dynamical running of the four-fermion
couplings that leads to a finite dynamical fermion mass, and hence an
infrared-meaningful theory in the continuum limit.  Now, returning to the 
phenomenon of magnetic catalysis, one is naturally led to ask: 
what are the effects of four-fermion interactions in the chiral symmetry 
breaking triggered by an external magnetic field?  A recent work by
Hong \cite{Hong} suggests that quantum effects of fermions in higher
Landau levels induce marginal four-fermion couplings (along with other
interactions).  In turn, the four-fermion interactions give rise to chiral
symmetry breaking when the external magnetic field is super strong.
Hong's analysis, though incomplete, is intriguing.  It hints at a
non-zero critical magnetic field in the magnetic catalysis of chiral
symmetry breaking.  The next question to ask is this:
Will a complete analysis of this phenomenon yield 
a non-zero critical gauge coupling as well as a
non-zero critical magnetic field?  Indications from the
analysis of strong-coupling QED are encouraging:
in strong-coupling QED, supplemented by the appropriate four-fermion
interactions, both the critical gauge coupling and the critical
four-fermion
coupling are non-zero.  This brings us back to the question posed at the
beginning of this talk: Does chiral symmetry breaking by an external field
shed light on the Pauli problem by giving a critical value of 
$\alpha$ close to 1/137?

\section*{Acknowledgements}  
This talk is based on 
works done in collaboration with 
C.N. Leung, D.-S. Lee, and A. Ackley,
supported in part 
by the U.S. Department of Energy under Grant No.
DE-FG05-85ER-40219 Task A.

\end{document}